\documentclass[prd,onecolumn]{revtex4}
\usepackage{dcolumn}
\usepackage{multirow}
\usepackage{graphicx}
\usepackage{amssymb}
\usepackage{bm}
\usepackage{hyperref}
\usepackage{epstopdf}
\usepackage{color}
\usepackage{mathrsfs}
\usepackage{amsmath,amssymb,amsthm}

\begin{document}
\title{Traversable holographic dark energy wormholes constrained by astronomical observations}
\author{Deng Wang}
\email{Cstar@mail.nankai.edu.cn}
\affiliation{Theoretical Physics Division, Chern Institute of Mathematics, Nankai University,
Tianjin 300071, China}
\author{Xin-he Meng}
\email{xhm@nankai.edu.cn}
\affiliation{{Department of Physics, Nankai University, Tianjin 300071, P.R.China}\\
{State Key Lab of Theoretical Physics,
Institute of Theoretical Physics, CAS, Beijing 100080, P.R.China}}
\begin{abstract}
In this letter, we investigate the traversable wormholes in the holographic dark energy (HDE) model constrained by the modern astronomical observations. First of all, we constrain the HDE model by adopting different data-sets, explore the cosmological background evolution of the HDE model, and find that the HDE model will be fitting better than the Ricci dark energy (RDE) model for the same SNe Ia data-sets by using the the so-called Akaike Information Criterions (AIC) and Bayesian Information Criterions (BIC) . Furthermore, we discover that if taking the SNe Ia data-sets, the wormholes will appear (open) when the redshift $z<0.027$. Subsequently, several specific traversable wormhole solutions are obtained, including the constant redshift function, traceless stress energy tensor, a special choice for the shape function as well as the case of isotropic pressure. Except for the first case, it is very necessary to theoretically construct the traversable wormholes by matching the exterior geometries to the interior geometries. Naturally, one can easily find that the dimensions of the wormholes for the left cases are substantially finite.

\end{abstract}
\maketitle

\section{Introduction}
In recent years, numerous and complementary observations have confirmed that the universe is undergoing a phase of accelerated expansion \cite{1,2,3}. Gradually mounting evidence of the cosmological expansion, coming from the measurements of Type Ia supernovae (SNe Ia), cosmic microwave background radiation (CMB), baryonic acoustic oscillations (BAO), observational Hubble parameter data (OHD) etc., indicates that the universe consists of some kind of negative enough pressure dubbed `` dark energy ''. The joint analysis of cosmological observations suggests that the universe is composed of about 73\% dark energy, 23\% dark matter, 4\% dust matter (baryons) and negligible radiation. Although we can affirm that the ultimate fate of the universe is determined by the characteristics of dark energy, so far, the nature of dark energy is still an enigma. For this reason, theorists have proposed many alternatives attempting to explore the origin of dark energy. At present, there appear to be two distinctive routines in which the universe could be made to accelerate:

$\bullet$ Physical dark energy models (PDE): the cosmological constant \cite{4}, phantom \cite{5}, quintessence \cite{6,7,8,9,10,11,12,13,14}, quintom \cite{15}, ghost condensates \cite{16,17}, Chaplygin gas (CG) \cite{18,19,20}, generalized Chaplygin gas (GCG) \cite{21,22}, bulk viscosity \cite{23,24,25,26,27,28}, decaying vacuum \cite{29}, Cardassian model \cite{29.1,30} etc.

$\bullet$ Geometrical dark energy models (GDE): f(R) gravity \cite{31,32,33}, Einstein-Aether gravity \cite{34,35}, scalar-tensor theories of gravity \cite{36,37,38,39,40,41,42}, braneworld models \cite{43,44,45,46}, Gauss-Bonnet gravity \cite{47,48,49,50}, Chern-Simons gravity \cite{51}, bimetric theories \cite{52,53,54}, Ho\v{r}ava-Lifschitz gravity \cite{55,56,57}, Galileon modification of gravity \cite{59}, Dvali-Gabadadze-Porrati gravity (DGP) \cite{59}, holographic dark energy (HDE) \cite{60,61,62,63}, Ricci dark energy (RDE) \cite{64,65,66,67} etc.

Although there are a lot of models and theories having been proposed to understand the nature of dark energy better, we still know little about dark energy today, and there appear to be a high degeneracy between the PDE models and the GDE models as before. Therefore, one may expect to solve this mysterious problem by developing a complete quantum gravity theory (QGT), but, unfortunately, that may be probably a rather difficult problem. Although a complete QGT has not been developed, we could still explore partly the nature of the dark energy by using the holographic principle \cite{68,69} which acts as an important result of present QGT (or string theory) for gravity phenomena. It is well known that the holographic principle is inspired by the investigations of quantum properties of black holes and shed some light on the cosmological problem and the dark energy problem. Generally speaking, for a quantum gravity system, a conventional quantum field theory includes a great deal of degrees of freedom, which will lead to the formation of black holes in order to break down the effectiveness of the quantum field theory in Minkowski background.

Because of the extraordinary thermodynamics of a black hole, Bekenstein \cite{70,71} proposed the `` maximal entropy postulation '' that, for an effective field theory with UV cutoff $\Lambda$ in a box of volume $L^3$, the maximal entropy behaves non-extensively, growing only as the surface area of the box, namely, there exists a famous Bekenstein entropy bound, $(L\Lambda)^3\leq S_{BH}=\pi L^2M^2_{pl} $, where $S_{BH}$ is the entropy of a black hole of radius $L$ which act as an IR cutoff and $M_{pl}=1/\sqrt{8\pi G}$ represents the reduced planck mass. The non-extensive scaling indicates that the conventional quantum field theory which supports the extensive scaling breaks down in a large volume. To alleviate this intractable problem, Cohen et al. \cite{72} proposed a stricter limitation, i.e., the energy bound, $L^3\Lambda^4\leq LM_{pl}^2$, which implies the total energy of the whole system in a region of given size $L$ should not exceed the mass of a black hole of the same size. Hence, one can obtain the conclusion that the maximal entropy is in order of $S_{BH}^{3/4}$. Based on this assumption, Li proposed the HDE model as follows
\begin{equation}
\rho_H=3c^2M^2_PL^{-2},
\end{equation}
where $\rho_H$ is holographic dark energy density and $c$ is a numerical factor. Since the HDE with Hubble horizon as its IR cutoff does not explain the accelerated mechanism of the universe, namely, the equation of state parameter is greater than $-1/3$ \cite{73}, Li \cite{74} suggested that the future event horizon instead of Hubble horizon could be used as the characteristic length:
\begin{equation}
R_{eh}(a)=a\int^\infty_t\frac{d\tilde{t}}{a(\tilde{t})}=a\int^\infty_a\frac{d\tilde{a}}{H\tilde{a}^2}.
\end{equation}
Furthermore, researchers find that this model gives an accelerating universe and is well compatible with current observations \cite{75,76,77}. It is obvious that the parameter $c$ plays a crucial role in the aforementioned HDE model, and we will give a more specific description that how the features of the HDE model depend on the values of parameter $c$ in the next section.

In this situation, we pay mainly our attention to investigate the astrophysical scale properties (wormholes) of the HDE model and its dependence on the evolution of the universe, by assuming the dark energy fluids is permeated everywhere in the whole bulk. As our previous works \cite{78,79,80}, we believe strongly that the deeper and deeper researches of wormholes the extremely astrophysical objects together with white dwarfs, black holes, neutron stars etc., may provide a window for new physics. Thus, it is necessary to make a brief review about wormhole physics in the following context.

Wormholes can be defined simply as handles or tunnels in the spacetime topology connecting different universes or widely separated regions of our universe via a throat \cite{82}. We think that the attractive and puzzling objects (wormholes) are widely studied for a long history, which is mainly based on the following two reasons:

$\bullet$ Wormhole can be acted as an effective and powerful tool for a rapid interstellar travel and is often an important routine to construct the so-called time machines.

$\bullet$ Based on the elegant discovery that the universe is undergoing the cosmic acceleration, a gradually increasing attention to the subjects (wormholes) has arisen significantly in connection with the global cosmology scale discovery. Because of the violation of the null energy condition (NEC), namely, $T_{\mu\nu}k^{\mu}k^{\nu}>0$, and consequently all of other energy conditions, where $T_{\mu\nu}$ is the stress-energy tensor and $k^{\mu}$ any future directed null vector, an interesting and subtle overlap between the two seemingly separated subjects occurs. To be more precise, if we parameterize the dark energy through an equation of state $\omega=p/\rho$, where $p$ is the spatially homogeneous pressure and $\rho$ the energy density of dark energy, we will get the conclusion that if $\omega<-1$, the wormholes will appear (i.e., the NEC is violated). At the same time, it is worth noting that this is the key starting point of our work.

Recently, there are also two reasons, giving us the newer and stronger motivation to explore the wormhole physics further, as follows:

$\bullet$ Three earlier studies \cite{83,84,85} have verified the possible existence of wormholes in the outer regions of the galactic halo and in the central parts of the halo, respectively, based on Navarro-Frenk-White (NFW) density profile and the Universal Rotation Curve (URC) dark matter model \cite{86,87}. In particular, the second result is an important supplement to the first one, thereby confirming the possible existence of wormholes in most of the spiral galaxies.

$\bullet$ In \cite{80}, we introduce the modern astrophysical observations into wormhole physics, which seems to be the first try in the literature. Constraining the model parameters by the astronomical observations, for a concrete dark energy model, one can apparently find that in which stage of the evolution of the universe the wormholes could appear (open) and/or disappear (close), give a strong restriction to the parameter range, reduce the numbers of the wormholes and provide a new perspective for the wormhole research starting from the observational cosmology.

In addition, lately, Fabrizio et al. have already constructed the analytic self-gravitating multi-Skyrmonic configurations and a self-gravitating cloud of interacting Pions \cite{88,89,90,91}. The corresponding geometries reflects the Lorentzian traversable wormhole with NUT parameters. Furthermore, it is mentionable that they proposed a matter field (i.e., the low dynamics of Pions plus the negative cosmological constant) to support the formation of wormholes instead of the usual exotic matter, and it seems to be the first Lorentzian traversable wormhole constructed in 3+1 dimensional General Relativity (GR) minimally coupled to a physical source (namely, the Pions) in which the only ``exotic matter'' needed to support it is a negative cosmological constant.

In the present letter, we intend to investigate the HDE traversable wormholes (belonging to the second class in \cite{80}) constrained by the modern astrophysical observations. To the best of our knowledge, wormhole geometries in the HDE model are never considered in the literature. Based on the new technique in paper \cite{80}, we expect to provide a systematic and detailed study for the HDE traversable wormholes.

This paper is organized in the following manner: In the next section, the HDE model is exhibited in order to be constrained by observations in the following context. In Section III, we constrain the HDE model by the SNe Ia, OHD, BAO and CMB data-sets. Moreover, we also discriminate the HDE model and RDE model in theoretical statistics by using the Akaike Information Criterions (AIC) and Bayesian Information Criterions (BIC). In Section IV, we investigate several traversable wormholes and the related properties and features, containing a specific choice for the redshift function, a special choice for the shape function, the traceless stress energy tensor case and the isotropic pressure case. In Section V, we make a discussion and conclude the present paper (we will take the units $c=\hbar=8\pi G=1$ throughout the context).

\section{The HDE model}
In this section, we will make a brief review about the HDE model. Considering the spatially flat Friedmann-Robertson-Walker (FRW) universe with the HDE component $\rho_{\Lambda}$ and the matter component $\rho_m$, so the first Friedmann equation can be written as
\begin{equation}
3M^2_{pl}H^2=\rho_{\Lambda}+\rho_m.
\end{equation}
Thus, the dimensionless Hubble parameter is
\begin{equation}
E(z)\equiv\frac{H(z)}{H_0}=[\frac{\Omega_{m0}(1+z)^3}{1-\Omega_\Lambda}]^{\frac{1}{2}},
\end{equation}
where $1+z=\frac{1}{a}$ represents the relation between the redshift and the scale factor. It is noteworthy that we will assume the spatial flatness and set $a_0=1$ in the whole context, where the subscript 0 represents the present-day value. In combination with Eq. (1) and the definition of the future event horizon Eq. (2), one can derive
\begin{equation}
\int^\infty_a\frac{d\ln\tilde{a}}{H\tilde{a}}=\frac{c}{Ha\sqrt{\Omega_\Lambda}}.
\end{equation}
At the same time, we notice that the first Friedmann equation Eq. (3) indicates
\begin{equation}
\frac{1}{Ha}=\sqrt{a(1-\Omega_\Lambda)}\frac{1}{H_0\sqrt{\Omega_{m0}}}.
\end{equation}
Then, substituting Eq. (6) into Eq. (5) and taking the derivative with respect to $z$ on both sides, one can easily obtain the dynamically differential equation of the fractional density of dark energy as follows:
\begin{equation}
\Omega_\Lambda'=-\frac{\Omega_\Lambda(1-\Omega_\Lambda)}{1+z}(1+\frac{2}{c}\sqrt{\Omega_\Lambda}),
\end{equation}
where the prime denotes the derivative with respect to $z$. One could obviously see that $c$ is the only parameter determining the dynamical behavior of the HDE. As a matter of fact, one could obtain the equation of state of the HDE by using the equation of energy conservation as follows:
\begin{equation}
\omega=-\frac{1}{3}(1+\frac{2}{c}\sqrt{\Omega_\Lambda}).
\end{equation}
It is easy to see that, if one sets $c<1$, the equation of state parameter will cross the phantom barrier $\omega=-1$ (or phantom divide), exhibiting an interesting behavior of a quintom-like model. If $c=1$, the evolutional behavior of the HDE model will be more and more like the standard cosmological model with the expansion of the universe, so as to the universe will tend ultimately to be a de Sitter universe. If $c>1$, the model will be a quintessence-like one all the time, i.e., $\omega>-1$, which naturally avoids approaching the big rip phase and tending to become a de Sitter universe ultimately. In addition, it has been shown that the HDE model exhibits a quintom-like behavior basically within one sigma error in the previous analysis of observation constraints \cite{75,76,77}.
\section{Constrain the HDE model by observations}
\subsection{Type Ia Supernovae}
The observations of SNe Ia have provided an forceful and effective tool to explore the expansion history of the universe. As is known to all, the SNe Ia observations directly measure the apparent magnitude $m$ of a supernova and its redshift $z$. Furthermore, one could define the distance modulus:
\begin{equation}
\mu_{th}(z_i)=m-M=5\log_{10}D_L(z_i)+\mu_0,
\end{equation}
where $M$ is the absolute magnitude which is believed to be a constant for all the SNes Ia, and $\mu_0=42.39-5\log_{10}h$, $h$ is the dimensionless Hubble parameter today in units of 100 $km^{-1}s^{-1}Mpc$. Subsequently, the luminosity distance redshift relation can be expressed as
\begin{equation}
d_L(z)=(1+z)\int^z_0\frac{dz'}{E(z';\theta)},
\end{equation}
where $\theta$ denotes the model parameters. In order to make constraints on the HDE model, we adopt the theoretical $\chi^2$ statistics for the parameter pair ($c$, $\Omega_{m0}$). The corresponding $\chi^2_S$ function for the SNe Ia analysis is
\begin{equation}
\chi^2_S=\sum^{580}_{i=1}[\frac{\mu_{obs}(z_i)-\mu_{th}(z_i;\theta)}{\sigma_i}]^2,
\end{equation}
where $\mu_{obs}(z_i)$ is the observed value of distance modulus for every supernovae, and $\sigma_i$ the corresponding $1\sigma$ error. According to \cite{92}, the minimization with respect to $\mu_0$ can be obtained by Taylor expanding $\chi^2_S$,
\begin{equation}
\chi^2_S=A-2B\mu_0+C\mu_0^2,
\end{equation}
where
\begin{equation}
A(\theta)=\sum^{580}_{i=1}[\frac{\mu_{obs}(z_i)-\mu_{th}(z_i;\theta;\mu_0=0)}{\sigma_i}]^2,
\end{equation}
\begin{equation}
B(\theta)=\sum^{580}_{i=1}\frac{\mu_{obs}(z_i)-\mu_{th}(z_i;\theta;\mu_0=0)}{\sigma_i^2},
\end{equation}
\begin{equation}
C=\sum^{580}_{i=1}\frac{1}{\sigma_i^2}.
\end{equation}
Hence, it is easy to find that $\chi^2_S$ is minimized when $\mu_0=\frac{B}{C}$ by calculating the transformed $\tilde{\chi}^2_{S}$:
\begin{equation}
\tilde{\chi}^2_{S}(\theta)=A(\theta)-\frac{[B(\theta)]^2}{C}.
\end{equation}
One can place constraints on the HDE model by adopting $\tilde{\chi}^2_{S}$ which is independent of $\mu_0$ instead of $\chi^2_S$.
\subsection{Cosmic Microwave Background and Baryonic Acoustic Oscillations}
As the important and effective supplements, we will adopt the CMB shift parameter and BAO to calculate the joint analysis in order to make the constraints more strictly. The CMB shift parameter $\mathcal{R}$ which may be the least independent model parameter that could be extracted from the CMB data-sets, is defined in paper \cite{93} as
\begin{equation}
\mathcal{R}=\sqrt{\Omega_{m0}}\int^{z_C}_0\frac{d\tilde{z}}{E(\tilde{z})},
\end{equation}
where $z_C$ is the redshift of recombination. The seven-year WMAP results \cite{94} have indicated the value of $z_C$ as $z_C=1091.3$ independent of the dark energy model and the shift parameter $\mathcal{R}=1.725\pm0.018$. The $\chi^2$ for the CMB observations can be defined as
\begin{equation}
\chi^2_{C}(\theta)=[\frac{\mathcal{R}(\theta)-1.725}{0.018}]^2.
\end{equation}
Another meaningful constraint comes from the large scale structure (LSS) data, and we adopt the measurements of the BAOs peak in the distribution of the Sloan Digital Sky Survey (SDSS) luminous red galaxies. In this situation, we just use $\mathcal{A}=0.469\pm0.017$ \cite{95} giving by the SDSS BAO measurement at $z_B=0.35$, where $\mathcal{A}$ is defined as
\begin{equation}
\mathcal{A}_{th}(\theta)=\sqrt{\Omega_{m0}}E(z_B)^{-\frac{1}{3}}[\frac{1}{z_B}\int^{z_B}_0\frac{d\tilde{z}}{E(\tilde{z})}]^{\frac{2}{3}}.
\end{equation}
Then, the $\chi^2$ for the BAO data-sets is given by
\begin{equation}
\chi^2_{B}=\sum^6_{i=1}[\frac{\mathcal{A}_{obs}(z_i)-\mathcal{A}_{th}(z_i;\theta)}{\sigma_{\mathcal{A}}}]^2,
\end{equation}
where $\sigma_{\mathcal{A}}$ denotes the statistical error one sigma and $\mathcal{A}_{obs}(z_i)$ the observed value of the distance parameter.
\subsection{Observational Hubble Parameter}
Generally speaking, there exist two main methods of independent observational $H(z)$ measurement, which are the `` differential age method '' and `` radial BAO method ''. More details can be found in papers \cite{96,97}, in which they summarize the updated OHD. As usual, the $\chi^2$ for the OHD can be defined as
\begin{equation}
\chi^2_{H}=\sum^{29}_{i=1}[\frac{H_0E(z_i)-H_{obs}(z_i)}{\sigma_i}]^2,
\end{equation}
where $H_{obs}(z_i)$ denotes the observed value of the OHD. Using the aforementioned trick, the minimization with respect to $H_0$ can be made by Taylor-expanding $\chi_{H}^2$ as
\begin{equation}
\chi^2_{H}(\theta)=AH_0^2-2BH_0+C,
\end{equation}
where
\begin{equation}
A=\sum^{29}_{i=1}\frac{E^2(z_i)}{\sigma_i^2},
\end{equation}
\begin{equation}
B=\sum^{29}_{i=1}\frac{E(z_i)H_{obs}(z_i)}{\sigma_i^2},
\end{equation}
\begin{equation}
C=\sum^{29}_{i=1}\frac{H^2_{obs}(z_i)}{\sigma_i^2}.
\end{equation}
Therefore, $\chi^2_H$ is minimized when $H_0=\frac{B}{A}$ by calculating the following transformed $\tilde{\chi}^2_{H}$ :
\begin{equation}
\tilde{\chi}^2_{H}=-\frac{B^2}{A}+C.
\end{equation}
\begin{figure}
\centering
\includegraphics[scale=0.5]{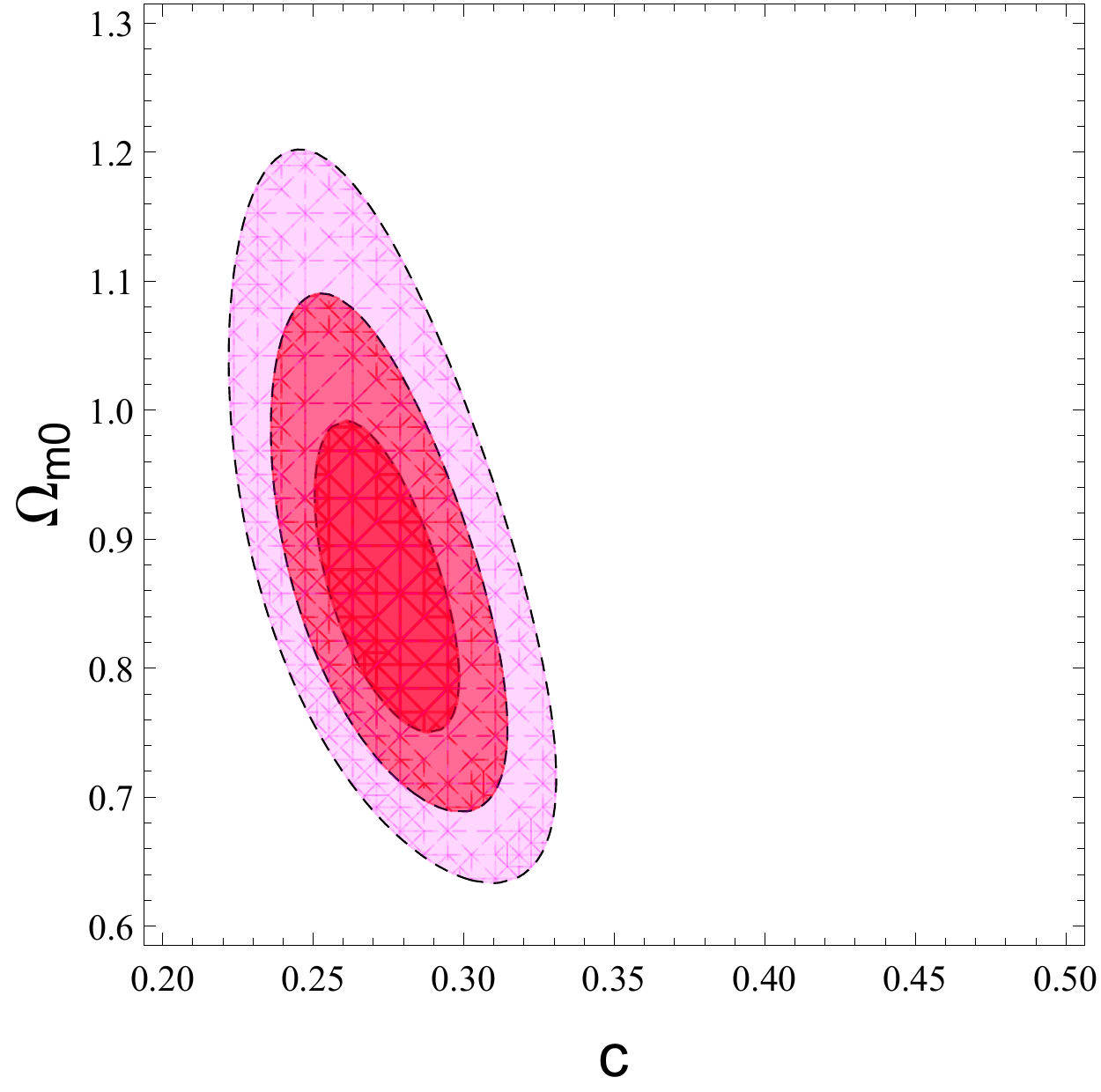}
\caption{1$\sigma$, 2$\sigma$ and 3$\sigma$ confidence ranges for parameter pair ($c$, $\Omega_{m0}$) of the HDE model, constrained by SNe Ia, BAO and OHD data-sets.}\label{1}
\end{figure}
\begin{figure}
\centering
\includegraphics[scale=0.5]{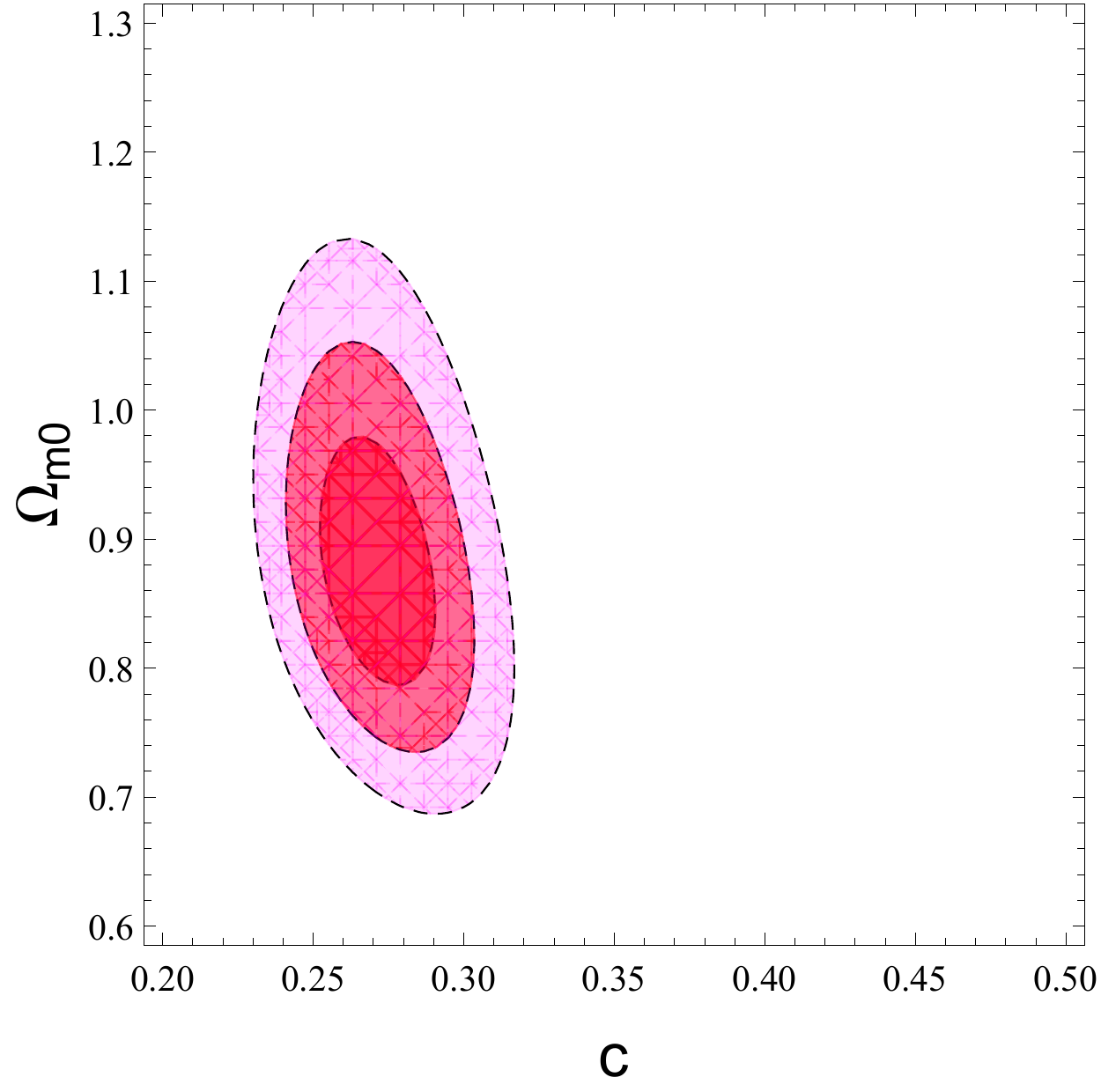}
\caption{1$\sigma$, 2$\sigma$ and 3$\sigma$ confidence ranges for parameter pair ($c$, $\Omega_{m0}$) of the HDE model, constrained by SNe Ia, BAO, CMB and OHD data-sets.}\label{2}
\end{figure}
\begin{figure}
\centering
\includegraphics[scale=0.5]{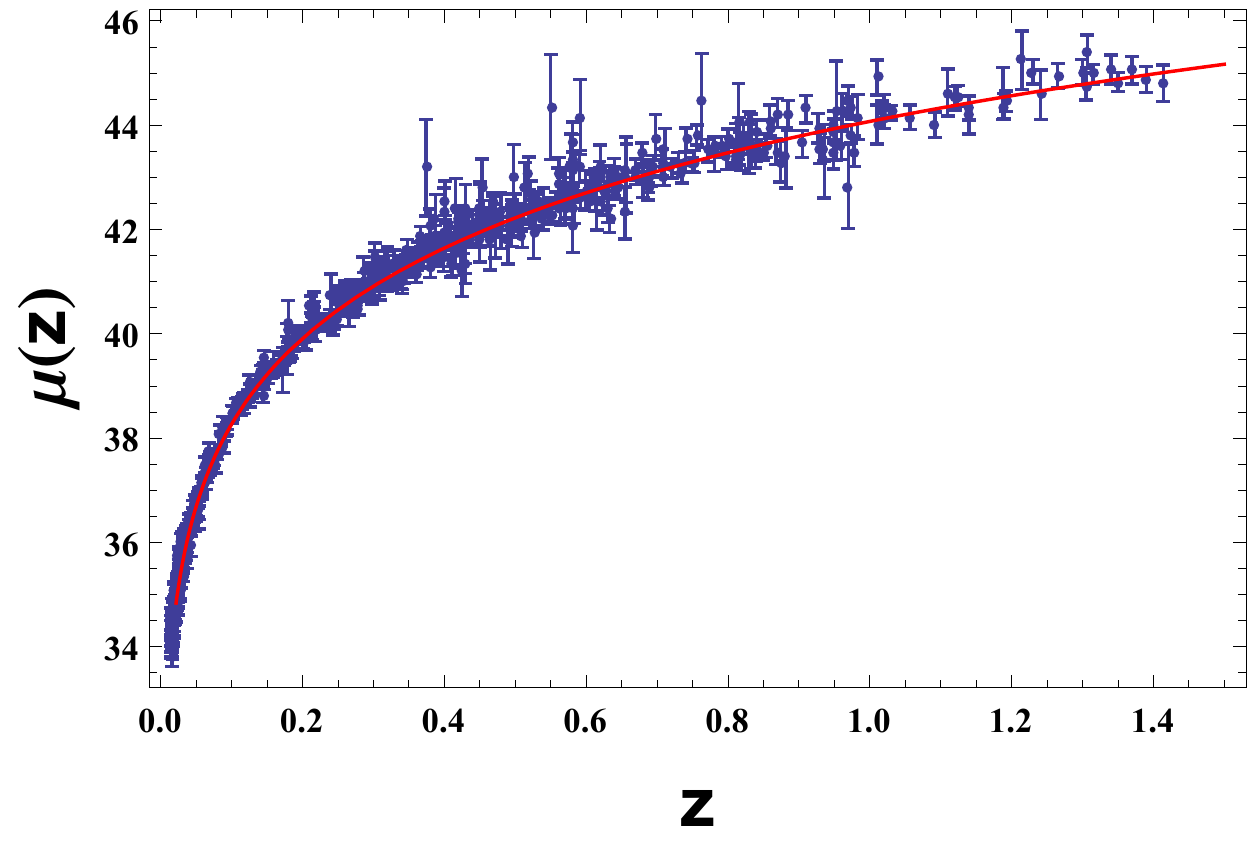}
\caption{The relation between the distance modulus and the redshift. The solid (red) line corresponds to the theoretical curve calculated from the model concerned. The dots with errors bar represent the 580 data points from the supernovae observations. Furthermore, one can find that the theoretical curve is well compatible with the observations.}\label{3}
\end{figure}
\begin{figure}
\centering
\includegraphics[scale=0.5]{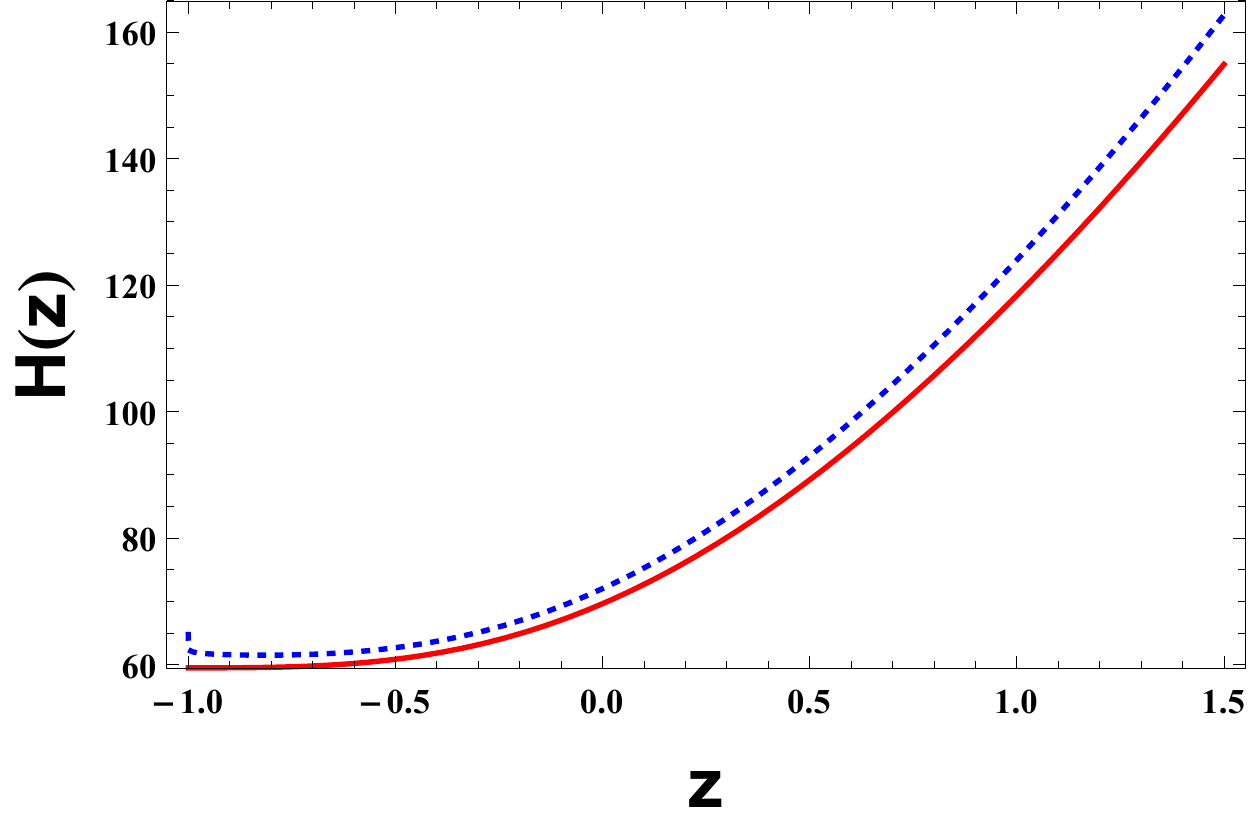}
\caption{The relation between the Hubble Parameter $H(z)$ and the redshift $z$. The solid (red) line and the dotted (blue) line correspond to the $\Lambda$CDM model and the HDE model, respectively.}\label{4}
\end{figure}
\begin{figure}
\centering
\includegraphics[scale=0.5]{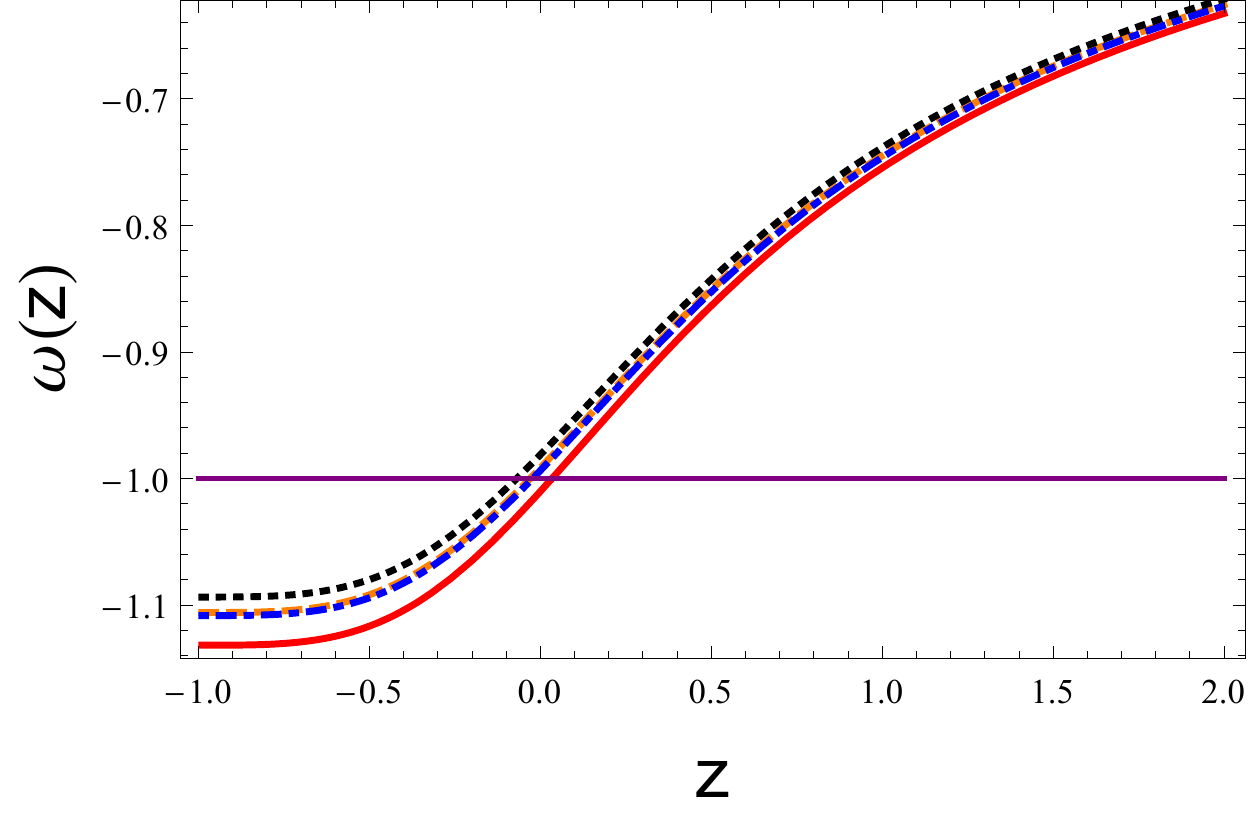}
\caption{The equation of state parameter $\omega(z)$ versus the redshift $z$ for the HDE model. The solid (red) line, the dashed (orange) line, the dot-dashed (blue) line as well as the dotted (black) line correspond to the curves of the best fitting values from SNe Ia alone, SNe Ia and OHD, SNe Ia, BAO and OHD as well as SNe Ia, BAO, CMB and OHD, respectively. The horizontal (purple) line corresponds to the $\Lambda$CDM model. In addition, one could easily find that the HDE model exhibit the evolution behavior of quintom-like model.}\label{5}
\end{figure}
One can conveniently place constraints on the HDE model by using $\tilde{\chi}^2_{O}$ which is independent of $H_0$ instead of $\chi^2_O$.
Subsequently, in the first place, we shall compute the combined constraints from SNe Ia, OHD and BAO data-sets, and the corresponding $\chi^2_1$ can be defined as
\begin{equation}
\chi^2_{1}=\tilde{\chi}^2_{S}+\chi^2_{B}+\tilde{\chi}^2_{H}.
\end{equation}
In the second place, we also calculate the joint constraints from SNe Ia, OHD, CMB and BAO data-sets. The corresponding $\chi^2_2$ can be defined as
\begin{equation}
\chi^2_{2}=\tilde{\chi}^2_{S}+\chi^2_{B}+\chi^2_C+\tilde{\chi}^2_{H}.
\end{equation}
\begin{table}[h!]
\begin{tabular}{ccccccc}
\hline
                      &SNe Ia           & SNe Ia+OHD     &SNe Ia+BAO+OHD   & SNe Ia+BAO+CMB+OHD&\\
\hline
$\chi^2_{min}$ & $562.225$ &$578.233$ &$578.232$ &$578.342$\\
$\Omega_{m0}$         & $0.280980$      &$0.273592$      & $0.274054$     & $0.270916$\\
$c$                   & $0.834997$       &$0.862902$      & $0.860212$     & $0.876781$ \\
\hline
\end{tabular}
\caption{The best fitting values of the model parameter pair ($c$, $\Omega_{m0}$) in the HDE model by using several different kinds of constraints: SNe Ia, BAO, CMB and OHD.}
\label{tab1}
\end{table}
The minimum values of the derived $\chi^2_{1}$ and $\chi^2_{2}$ and the best fitting values of the parameters are listed in Table. \ref{tab1}. The likelihood distributions of the parameters ($a$, $b$) in the two distinct constraints ($\chi^2_{1}$ and $\chi^2_{2}$) are exhibited in Fig. \ref{1} and Fig. \ref{2}, respectively. Furthermore, it is very useful and necessary to show the relation between the distance modulus and redshift (see Fig. \ref{3}), since one can get an apparent picture about the evolutional behavior of the universe in the HDE model through adopting the best fitting values of the model parameters (see Fig. \ref{4}). At the same time, one can find that the HDE model is well compatible with the astrophysical observations. Moreover, the equation of state parameter $\omega$ with respect to the redshift $z$ from four observational constraints including the SNe Ia alone, SNe Ia and OHD, SNe Ia, BAO and OHD as well as SNe Ia, BAO, CMB and OHD (see Table. \ref{tab1}), are exhibited in Fig. \ref{5}.

From Fig. \ref{3}, it is easy to be seen that the theoretical curve of distance modulus $\mu(z)$ with respect to redshift $z$ is well consistent with the 580 SNe Ia samples. In Fig. \ref{4}, one can obviously discover that the cosmological background evolution of the HDE model is also well compatible with the $\Lambda$CDM model at the present epoch. Nonetheless, in the far future and in the remote past, the discrepancies exist and finally, the universe will tent to be undergoing the phase of accelerated expansion in a higher velocity (but finite) than the standard cosmological model. Comparing with the RDE model \cite{80}, we find that HDE model will be more consistent with the $\Lambda$CDM model, although the RDE model is substantially compatible with the $\Lambda$CDM model in the past. From Fig. \ref{5}, one can not only get the evolution behavior of the universe in the HDE model (quintom-like), but also clearly discover that the change of the type of the cosmic matter (quintessence-like or phantom-like) by comparing with the $\Lambda$CDM model. In addition, one may find that the two data constraints have a high degeneracy (namely, SNe Ia and OHD as well as SNe Ia, BAO and OHD) since we just use one BAO node.
\subsection{Akaike Information Criterions and Bayesian Information Criterions}
After constraining the HDE model, we think that it is constructive to statistically compare the HDE model and the RDE model (see \cite{80}) by using the same SNe Ia data-sets. Therefore, we will adopt the so-called Akaike Information Criterions (AIC) \cite{98} and Bayesian Information Criterions (BIC) \cite{99} to discriminate the two dark energy models. Furthermore, the two criterions can be defined as
\begin{equation}
AIC=\chi^2+2n \qquad\qquad and \qquad\qquad BIC=\chi^2+n\ln N,
\end{equation}
where $n$ and $N$ denote the number of the model parameters and the number of the used data points.
Hereafter, adopting the same SNe Ia data constraint for the two models, one can get the following relationship:
\begin{equation}
A_H=566.225<A_R=568.907 \qquad\qquad and \qquad\qquad B_H=574.951<B_R=581.996,
\end{equation}
where $A_H$, $A_R$, $B_H$ and $B_R$ denote the AIC value for the HDE model, AIC value for the RDE model, BIC value for the HDE model and BIC value for the RDE model, respectively. Generally speaking, in theoretical statistics, the smaller the AIC and BIC values are for different models constrained by the same data samples, the model will be fitting better for the present data-sets. Hence, it is not difficult to discover that the HDE model will be better than the RDE model for the same SNe Ia data samples, which verify the same conclusion obtained from the Fig. \ref{4} once again.

As mentioned above, we would like to explore the wormholes in the HDE model by astronomical observations. Particularly, one can discover that wormhole spacetime configurations will appear at $z<0.027$ and $z<-0.06086$ which comes from the SNe Ia constraint and the joint constraint of SNe Ia, BAO, CMB and OHD, respectively. More specifically, one can obtain more valuable information about the wormhole physic for a concrete cosmological model. For instance, we can apparently discover that in which stage of the evolution of the universe the wormholes can appear (open) and/or disappear (close), place a strong restriction to the parameter range, reduce the numbers of the wormholes and provide a new perspective for the wormhole research from the observational cosmology.

\section{Traversable HDE wormholes}
\subsection{The Basic Equations}
Consider the wormhole geometry given by the following static and spherically symmetric metric
\begin{equation}
ds^2=-e^{2\Phi(r)}dt^2+\frac{dr^2}{1-\frac{b(r)}{r}}+r^2(d\theta^2+\sin^2\theta d\phi^2),
\end{equation}
where $b(r)$ and $\Phi(r)$ are arbitrary functions of the radial coordinate $r$, denoted as the shape function and redshift function, respectively \cite{82}. It is worth noting that the radial coordinate $r$ is non-monotonic in order that it can decrease from the infinity to the minimum value $r_0$, represents the radius of the wormhole throat, where $b(r_0)=r_0$.

To form a traversable wormhole, as expressed in \cite{80,82}, in general, there are four fundamental ingredients as follows:

$\bullet$ The most fundamental requirement to form a wormhole is violating the NEC, i.e., $T_{\mu\nu}k^{\mu}k^{\nu}>0$.

$\bullet$ Satisfy the so-called flaring out conditions that can be expressed as: $b(r_0)=r_0$, $b'(r_0)<1$ and $b(r)<r$ when $r>r_0$.

$\bullet$ $\Phi(r)$ must be finite everywhere, in order to avoid an horizon, which can be identified the surfaces with $e^{2\Phi(r)}\rightarrow0$.

$\bullet$ The asymptotically flatness must be satisfied, which demands that $b/r\rightarrow0$ and $\Phi\rightarrow0$ when $r\rightarrow\infty$. As a matter of fact, one could not obtain directly an asymptotically flat wormhole solution by solving the Einstein Field Equations (EFE) for a concrete cosmological model. However, one can construct an asymptotically flat wormhole spacetime, by matching an exterior flat geometry (such as the Schwazschild geometry and Reissner-Norsdtr\"{o}m geometry) to the interior geometry at a junction radius $a$.

By using the EFE, namely, $G_{\mu\nu}=T_{\mu\nu}$, one can obtain the corresponding relationships as follows:
\begin{equation}
b'=r^2\rho,
\end{equation}
\begin{equation}
\Phi'=\frac{b+r^3p_r}{2r^2(1-b/r)},
\end{equation}
where the prime denotes a derivative with respect to $r$, $\rho(r)$ is the matter energy density and $p_r(r)$ is the radial pressure of dark energy fluid. At the same time, one can also derive from the conservation equation of the stress-energy tensor $T^{\mu\nu}_{\hspace{3mm};\nu} = 0$ with $\mu=r$ that
\begin{equation}
p'_r=\frac{2}{r}(p_t-p_r)-(\rho+p_r)\Phi',
\end{equation}
where $p_t(r)$ represents the transverse pressure measured in the orthogonal direction to radial direction. The above equation could also be interpreted as the relativistic Euler equation or the hydrostatic equation for equilibrium for the material threading a wormhole.

Form Eq. (8), one can equivalently derive the equation of state of the HDE model
\begin{equation}
p=-\frac{1}{3}(1+\frac{2}{c}\sqrt{\Omega_\Lambda})\rho.
\end{equation}
For simplicity, we denote $A=\frac{1}{3}(1+\frac{2}{c}\sqrt{\Omega_\Lambda})$ hereafter. In addition, we must point out that the pressure in HDE equation of state represents the radial pressure, thus, Eq. (33) can be rewritten as
\begin{equation}
p_r=-A\rho.
\end{equation}
At first glance, this equation of state seems to be the one in $\omega$CDM cosmology. Nonetheless, it is noteworthy that the newly redefined parameter $A$ contains two model parameters $c$ and $\Omega_\Lambda$, which is substantially important in the following contents. Using Eqs. (30-31), one can obtain that
\begin{equation}
\Phi'(r)=\frac{b-Arb'}{2r^2(1-\frac{b}{r})}.
\end{equation}
Subsequently, through using the condition $b'(r_0)<1$ in the HDE equation of state, evaluated at the throat radius $r=r_0$, we demonstrate that the energy density at $r_0$ is $\rho(r_0)=\frac{1}{Ar_0^2}$. Furthermore, we can obtain the relationship combining Eq. (30) and $b'(r_0)<1$ as follows
\begin{equation}
A>1.
\end{equation}
It is not difficult to find that the relationship is the same with that in $\omega$CDM cosmology, since the equations of state of these models all belong to the perfect equation of state. However, entirely distinct theoretical motivations are depicted in these different models. Also, the same relationship can be obtained from the violation of NEC, evaluated at the wormhole throat, i.e., $p_r(r_0)+\rho(r_0)<0$.
\section{Specific solutions}
\subsection{Constant Redshift function}
For a constant redshift function $\Phi=C$ (the most useful and the simplest case), one can obtain the shape function in the following manner:
\begin{equation}
b(r)=r_0(\frac{r}{r_0})^{\frac{1}{A}}.
\end{equation}
It is not difficult to be checked that $b(r)<r$ which satisfies the flare out condition when $r>r_0$. Evaluating at the throat $r_0$, one can derive
\begin{equation}
b'(r_0)=\frac{1}{A}.
\end{equation}
Subsequently, if adopting the best fitting values of the parameters in Table. \ref{tab1} from the mentioned-above joint constraints of SNe Ia, BAO, CMB and OHD data-sets, one can demonstrate $b'(r_0)=0.989764<1$. Intriguingly, the wormhole solution is not only asymptotically flat but also traversable, since $\Phi$ is finite everywhere and $b/r\rightarrow0$ when $r\rightarrow\infty$. Hence, the dimensions of the wormhole may be considerably large in principle.

According to \cite{100}, one can also consider an obvious relation between the transverse pressure and the energy density, namely, $p_t=\alpha\rho$, so we can obtain from replacing it in Eq. (32):
\begin{equation}
A\rho'=\frac{2}{r}(A+1)\rho.
\end{equation}
By using Eq. (30), this equation can be solved analytically, and we get $\alpha=\frac{1-5A}{2}<0$. Thus, the lateral pressure can be rewritten as
\begin{equation}
p_t=\frac{1-5A}{2}\rho.
\end{equation}
In connection with inequality (36), one could have the conclusion that $p_t\approx-2\rho<p_r$. Furthermore, if continuing using the best fitting values of the parameters from the aforementioned joint analysis, one can obtain $p_t=-2.02586\rho$. Therefore, we can see that the astrophysical observations provide a more precise and more physical description for the wormhole research.

Note that the most interesting consideration in wormhole physics may be to explore the traversability of a wormhole configuration. For this purpose, we will use the formulas in paper \cite{80} to derive the necessary condition as follows:
\begin{equation}
v\leq r_0\sqrt{\frac{Ag_\oplus}{(A-1)}},
\end{equation}
where $v$ denotes the traversal velocity and $g_\oplus$ 1 Earth's gravitational acceleration. It is noteworthy that we have assumed the height for a traveler to be 2 m. Then, if setting $r_0=100$ m and considering the best fitting value $(0.834997, 0.280980)$ from the SNe Ia analysis, one can obtain the velocity $v\approx3094.18$ m/s.
Subsequently, if continuing to consider the junction radius is given by $a=10000$ m, one can also get $\Delta\tau\approx\Delta t\approx6.46376$ s the traversal times,
according to $\Delta\tau\approx\Delta t\approx2a/v$ \cite{80,101,102}.

\subsection{The Traceless Stress Energy Tensor}
Consider the interesting case of the traceless stress energy tensor, which is always associated to the so-called Casimir effect with a massless field. It is worth noting that, sometimes, the Casimir effect can be theoretically invoked to provide the exotic matter (i.e., NEC violating matter) to the system considered at hand. Hence, using the traceless stress energy tensor, $T=-\rho+p_r+2p_t=0$, one can get the following equaiton
\begin{equation}
2(1-\frac{b}{r})[\Phi''+\Phi'+\frac{2\Phi'}{r}-\frac{b'r-b}{2r(r-b)}\Phi'-\frac{b'r-b}{2r^2(r-b)}]-\frac{b'}{r^2}-\frac{b}{r^3}=0.
\end{equation}
In principle, one can solve this differential equation exactly by inserting a special shape function or a specific redshift function. As a matter of fact, one can easily find that inserting a redshift function will be easier than imposing a specific shape function.

For instance, taking into account the case of $\Phi=C$, interestingly, one can discover that the shape function can be arbitrary function of the radial coordinate $r$. Thus, in the case of extremely physical condition, the shape function and/or redshift function also will exhibit some novelty properties and characteristics, which has been beyond the scope of the present letter. Subsequently, making an appropriate choice $\Phi=\ln(\frac{r}{r_0})$ for the redshift function, it follows that
\begin{equation}
2(1-\frac{b}{r})[\frac{2}{r^2}-\frac{b'r-b}{r^2(r-b)}]-\frac{b'r+b}{r^3}=0.
\end{equation}
Solving this equation, one can get
\begin{equation}
b(r)=\frac{1}{3}(2r+\frac{r_0^2}{r}).
\end{equation}
It is easy to be checked that the shape function satisfies the flare out condition. Unfortunately, the solution represents a non-asymptotically flat spacetime (i.e., $b(r)/r\rightarrow2/3\nrightarrow0$ and $\Phi\rightarrow\infty$ when $r\rightarrow\infty$). However, as mentioned above, one can theoretically construct a traversable wormhole through gluing an exterior geometry into an interior geometry.
\subsection{$b(r)=r_0+\frac{1}{A}(r-r_0)$}
Consider a specific shape function $b(r)=r_0+\frac{1}{A}(r-r_0)$, which is analogous to the choice in \cite{78}. Using Eq. (35), one can obtain
\begin{equation}
\Phi'(r)=-\frac{1}{2r},
\end{equation}
it follows that
\begin{equation}
\Phi(r)=C_1-\frac{1}{2}\ln r,
\end{equation}
where $C_1$ is an arbitrary integration constant. At the same time, it is easy to demonstrate that this solution is not asymptotically flat, consequently and non-traversable in the relatively large region. However, we can also construct a traversable one by matching the exterior geometry into the interior spacetime geometry at a junction radius $d$. Additionally, the constant $C_1$ is given by
\begin{equation}
C_1=\Phi(a)+\frac{1}{2}\ln(\frac{a}{r}).
\end{equation}
Now this solution represents a traversable wormhole since the redshift function is finite in the small range $d\geq r\geq r_0$ by a cutoff of the stress energy tensor.

Hereafter we will adopt the the so-called method of `` volume integral quantifier '' (VIQ), which has been widely used in the past ten years to quantify the total amounts of the exotic matter by calculating the definite integrals $\int T_{\mu\nu}U^\mu U^\nu dV$ and $\int T_{\mu\nu}k^\mu k^\nu dV$, to analyze the HDE model. Note that the amounts of the exotic matter can be defined as how negative the values of the integrals become. Furthermore, using the quantity $I_V=\int[p_r(r)+\rho]dV$ (based on the NEC) for the wormhole in the small range, one can get
\begin{equation}
I_V=[(r-b)\ln(\frac{e^{2\Phi}}{1-\frac{b}{r}})]^d_{r_0}-\int^d_{r_0}(1-b')[\ln(\frac{e^{2\Phi}}{1-\frac{b}{r}})]dr
=\int^d_{r_0}(r-b)[\ln(\frac{e^{2\Phi}}{1-\frac{b}{r}})]'dr.
\end{equation}
It is noteworthy that the first boundary term can vanish by considering the asymptotical flatness. Then, we can obtain the aforementioned definitive integral as follows
\begin{equation}
I_V=(\frac{1}{A}-1)(d-r_0).
\end{equation}
Subsequently, if we adopt the best fitting values of the model parameters from the constraint of the SNe Ia data-sets alone, the mentioned-above equation can be rewritten as
\begin{equation}
I_V=-0.0102361(d-r_0).
\end{equation}
It is not difficult to verify that the integral will approach zero when taking the limit $d\rightarrow r_0$, i.e., $I_V\rightarrow0$. Furthermore, this implies that one can theoretically construct a traversable wormhole with infinitesimal amounts of ANEC violating HDE matter. Besides, one can discover that this useful method may provide more information about the total amount of ANEC violating matter in the global spacetime \cite{82}.
\begin{figure}
\centering
\includegraphics[scale=0.7]{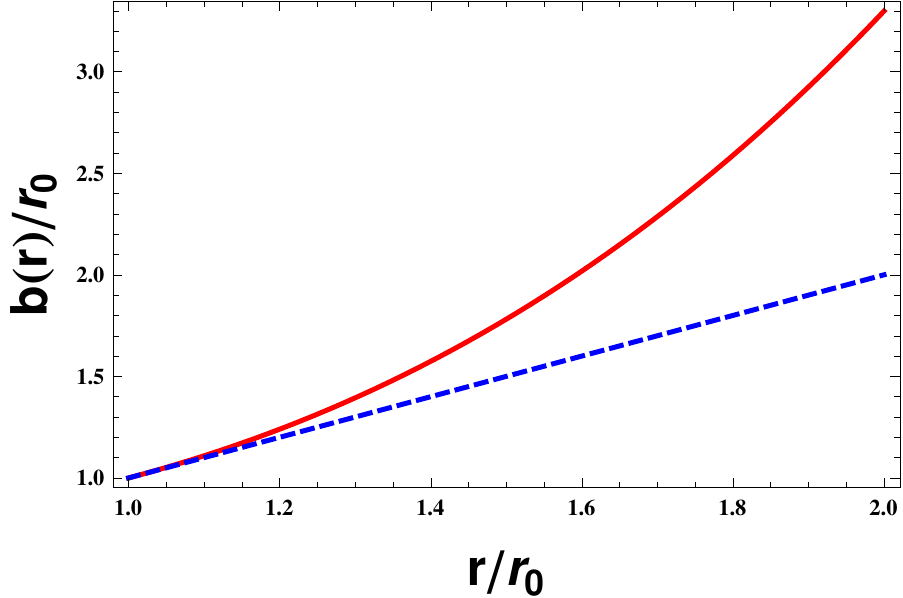}
\caption{To form a wormhole configuration, due to one of the flare out conditions $b(r)<r$, one could find in principle the wormholes dimension can be arbitrarily large. As a matter of fact, since the redshift function $\Phi(r)=\ln(\frac{r}{r_0})$ does not satisfy the flatness condition, the wormhole dimension is still substantially finite. The dashed (blue) line corresponds to the line of $b(r)=r$, the solid (blue) line represents the line of $b(r)$ for the case of isotropic pressure}\label{6}
\end{figure}
\subsection{The Isotropic Pressure}
Using Eq. (32) and taking into account the specific case of isotropic pressure, namely, $p_r=p_t$, one can conveniently obtain the following relationship
\begin{equation}
A\rho'=(1-A)\rho\Phi'.
\end{equation}
After rearranging terms and integrating on both sides, it follows that
\begin{equation}
\rho(r)=C_2e^{\frac{1-A}{A}\Phi(r)},
\end{equation}
where $C_2$ denotes an integration constant. Take note of $\rho(r_0)=1/Ar_0^2$, one can get
\begin{equation}
C_2=\frac{1}{Ar_0^2}e^{\frac{A-1}{A}\Phi(r_0)}.
\end{equation}
Subsequently, setting the redshift function as $\Phi(r)=\ln(\frac{r}{r_0})$ and substituting Eq. (52) into Eq. (30), one can obtain the following shape function \begin{equation}
b(r)=\frac{r(\frac{r}{r_0})^{1+\frac{1}{A}}+2Ar_0}{2A+1}.
\end{equation}
One could easily find that this solution satisfies the flare out conditions and reflects a non-asymptotically flat spacetime configuration. At the same time, we also can obtain the conclusion that the function $f(r)=r-b(r)$ in this case is monotonically increasing in the range $r\in(r_0, \infty)$ (see Fig. \ref{6}), which means that the dimensions of the wormhole geometry can be arbitrarily large. However, the redshift function does not satisfy the flatness condition so as to one can construct a traversable wormhole in a finite region by pasting an exterior geometry onto the interior geometry. Therefore, one obtain the same conclusion in this case for the HDE model with that for the RDE model (i.e., finite wormhole dimensions).
\section{Concluding Remarks}
The wonderful and elegant discovery that the universe is undergoing an phase of accelerated expansion, has given us an important chance to explore the wormhole physics further. To be more specific, one can regard the dark energy fluid as the exotic matter, and find the corresponding phantom matter to form a wormhole in any cosmological model. Although the nature of the dark energy is still to be determined in the future, we can provide a quantitative description for the wormhole spacetime configuration by introducing the modern astrophysical observations into the field of wormholes. This step means that one can apparently find that in which stage of the evolution of the universe the wormholes may appear (open) and/or disappear (close), give a strong restriction to the parameter range, avoid the arbitrarily mathematical choice for the model parameters and provide a new perspective for the wormhole research from the observational cosmology.

In this letter, we have studied the traversable wormholes constrained by the different data-sets in the HDE model. At first, through data fitting, we find the best fitting values of the parameter pair $(c, \Omega_{m0})$, make the contour plots for two joint constraints, explore the cosmological background evolution of the HDE model ,and discover that the HDE model will better consistent with the $\Lambda$CDM model than the RDE model. Furthermore, we can obtain the similar conclusion that the HDE model will be better compatible with the SNe Ia observations than the RDE model by using the so-called AIC and BIC. Subsequently, since we have found that the wormhole configurations will appear (open) when $z<0.027$, four specific solutions are analyzed vividly. For the first case of constant redshift function, we discover that the solution represents one both asymptotically flat and traversable wormhole, and explore the travsabilities of this wormhole. In the second case of traceless stress energy tensor, we construct a traversable wormhole in principle. For the third case of the specific shape function, we have constructed a traversable wormhole with infinitesimal amounts of ANEC violating HDE matter by using the so-called VIQ. It is worth noting that this method may provide more information about the total amount of ANEC violating matter in the whole spacetime. In the last case of isotropic pressure, we also theoretically construct a traversable wormhole, but the dimension of this wormhole is very finite.

Our coming work could be to take into consideration the dynamics of the wormhole spacetime, investigate the relationship between the energy conditions and wormhole configurations, and expect to constrain more cosmological models containing the HDE model by more accurate observations.
\section{acknowledgements}
This study is supported in part by the National Science Foundation of China. The authors would like to thank Prof. Jing-Ling Chen for helpful comments and discussions, and Guang Yang and Sheng-Sen Lu for programming. During the present work prepared period, we are grateful to Professors Bharat Ratra and Saibal Ray for very interesting communications on gravitational waves physics in cosmology and compact star formation as well as wormhole astrophysics.

\end{document}